\documentclass[twocolumn,twoside,dvips,showpacs,showkeys,preprintnumbers,aps,pre,amsmath,amssymb]{revtex4}

\usepackage{graphicx}
\usepackage{bm}
 \usepackage[usenames]{color}
\usepackage{color}
\definecolor{RED}{rgb}{1,0,0}
\usepackage{hyperref}

\def\hb2#1{{\color{black}{#1}}}

\renewcommand{\vec}[1]{\mathbf{#1}}

\newcommand{\Eq}[1]{Eq.~(\ref{eq:#1})}

\newcommand{\Fig}[1]{Fig.~\ref{fig:#1}}

\newcommand{\sect}[1]{\ref{sec:#1}}
\newcommand{\Sec}[1]{Sect.~\sect{#1}}

\newcommand{\ie}{\emph{i.e.,}~}

\newcommand{\cf}{\emph{cf.}~}

\newcommand{\Fcapi}{{\ensuremath{F_c}}}

\newcommand{\vareps}{\ensuremath{\varepsilon}}

\begin{document}


\title{Arrest stress of uniformly sheared wet granular matter}

\author{S. H. Ebrahimnazhad Rahbari} \affiliation{Department of
  Physics, Plasma and Condensed Matter Computational Laboratory,
  Faculty of Sciences, Azarbaijan Shahid Madani University, 51745-406
  Tabriz, Iran}

\author{M. Brinkmann} \affiliation{Max-Planck-Institut f\"ur Dynamik
  und Selbstorganisation (MPI DS), 37077 G\"ottingen, Germany}
\affiliation{Experimental Physics, Saarland University, 66123 Saarbr{\"u}cken} 
  
\author{J. Vollmer} \affiliation{Max-Planck-Institut f\"ur Dynamik und
  Selbstorganisation (MPI DS), 37077 G\"ottingen, Germany}
\affiliation{Fakult\"at f\"ur Physik, Universit\"at G\"ottingen, 37077
  G\"ottingen, Germany}


\date{\today}

\begin{abstract}
  We conduct extensive independent numerical experiments considering
  frictionless disks without internal degrees of freedom (rotation
  etc.) in two dimensions. We report here that for a large range of
  the packing fractions below random-close packing, all components of
  the stress tensor of wet granular materials remain finite in the
  limit of zero shear rate. This is direct evidence for a
  fluid-to-solid arrest transition. The offset value of the shear
  stress characterizes plastic deformation of the arrested state
  {which corresponds to {\em dynamic yield stress} of the system}. {
    Based on an analytical line of argument, we propose that the mean
    number of capillary bridges per particle, $\nu$, follows a
    non-trivial dependence on the packing fraction, $\phi$, and the
    capillary energy, $\vareps$. Most noticeably, we show that $\nu$
    is a generic and universal quantity which does not depend on the
    driving protocol.} Using this universal quantity, we calculate the
  arrest stress, $\sigma_a$, analytically based on a balance of the
  energy injection rate due to the external force driving the flow and
  the dissipation rate accounting for the rupture of capillary
  bridges. The resulting prediction of $\sigma_a$ is a non-linear
  function of the packing fraction $\phi$, and the capillary energy
  $\vareps$. This formula provides an excellent, parameter-free
  prediction of the numerical data. Corrections to the theory for
  small and large packing fractions are connected to the emergence of
  shear bands and of contributions to the stress from repulsive
  particle interactions, respectively.

\end{abstract}



\maketitle

\section{Introduction}

Careful studies for dry granular materials suggest that the arrest of
granular flow is universal in the sense that {this non-equilibrium
  transition} admits a {continuum description}. This points to the
existence of an equation of state for dry granular
matter~\cite{gdrmidi2004, losert2013, cruz2005,
  Jop2006,delannay2007,forterre2008}.
Similarly, various features of phase transition in vertically agitated
wet granular materials can faithfully be described in terms of
thermodynamic
concepts~\cite{herminghaus2004,fingerle2008,huang2009,roeller2011,strauch2012}.

Recently, a number of papers
\cite{RahbariVollmerHerminghausBrinkmann2010,huang2012,may2013,RoellerBlaschkeHerminghausVollmer2014,rahbari2013}
also addressed phase transitions in sheared wet granular matter. In
particular, the arrest of wet granular flows driven by external forces
with a cosine profile was attributed to a crossover of the power
injected into the flow by the external field and the power dissipated
in the rupturing of capillary bridges
\cite{RahbariVollmerHerminghausBrinkmann2010,rahbari2013,RoellerBlaschkeHerminghausVollmer2014}.

This provided a quantitative description of the arrest of shear flows
of bidisperse disks \cite{RahbariVollmerHerminghausBrinkmann2010} and
dumbbells \cite{rahbari2013} with a fixed density, and of
three-dimensional gravity-confined flow of monodisperse grains
\cite{RoellerBlaschkeHerminghausVollmer2014}. The studies established
that the minimal external forcing required to maintain flow can be
calculated along the same line for transitions in fixed-density and
fixed-pressure settings. {Also, for both wet disks and dumbbells, a
  large hysteresis was identified in which the fluid-to-solid arrest
  stress, $\sigma_a$, or the dynamic yield stress, has been found to
  be smaller than the solid-to-fluid yield stress, or the static yield
  stress. This is in accord with the conventional wisdom where the
  stress at yield point is conceived to be larger than the minimum
  (plastic) stress which is required to maintain the flow. Moreover,
  in a recent study, the difference of the static and dynamic yield
  stress is pinned down to the nonmonotonicity of the flow
  curves. Furthermore, it was shown that the nonmonotonicity of the
  flow curves indicates a shear banding
  instability~\cite{IraniChaudhuriHeussinger2014}.}

\begin{figure}
  \centering
  \includegraphics[scale=0.255]{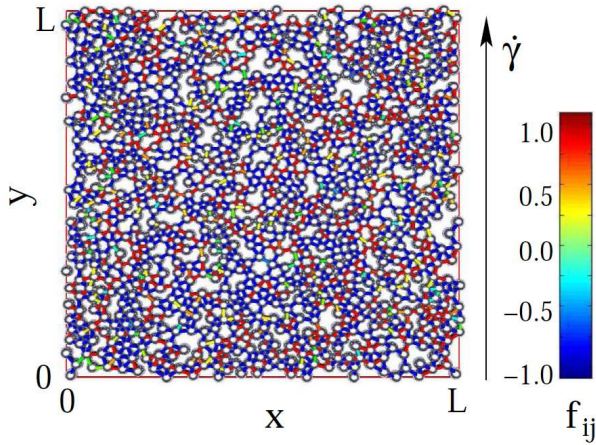}
  \caption{(color online) Setup of the simulations with a snapshot of
    the system.  It comprises of a square box of length $L$ sheared in
    vertical direction with a shear rate, $\dot\gamma$, and
    Lees-Edwards (skew periodic) boundary conditions.  Disks are
    indicated by circles that represent their actual size.  The
    strength of the forces, $f_{ij}$, acting between the {disks} $i$
    and $j$ is indicated by lines connecting the {disks}, where the
    color marks the strength of the force according to the color bar
    to the left. On the color palette, negative (positive) numbers
    correspond to attractive (repulsive) force. The snapshot shows
    a uniformly sheared state of a simulation of size, $L=30$, shear
    rate, $\dot\gamma=0.03$, and capillary-bridge energy,
    $\vareps=0.05$, {and packing density of disks, $\phi = 0.8$.}
  \label{fig:setting}}
\end{figure}

Here, we augment these studies by addressing the arrest stress in
systems with a prescribed global shear rate, $\dot{\gamma}$. Rather
than settings with complex flow profiles we consider for our present
study uniform shear flows in a Lees-Edwards~\cite{Lees1972, Allen1987}
periodic boundary flow geometry (\Fig{setting}).
This point of view is dual to the study of force-controlled systems
where we prescribe a force amplitude and measure the minimal force
required to sustain flow.
In the present study we will address the arrest stress, \ie the
smallest value of the stress observed when decreasing the shear shear
rate in spatially uniform granular flows. By definition of the stress
tensor, the product of this arrest stress and the system size amounts
to the minimal force that creeping flows with uniform shear profile
exert on the borders of a shear cell. This correspondence provides a
quantitative prediction of the arrest stress where the only adjustable
parameters have been measured in the force-controlled setting
\cite{RahbariVollmerHerminghausBrinkmann2010}.  The agreement provides
further support for the modeling of the flow threshold based on energy
dissipation arguments that {were established} in
\cite{RahbariVollmerHerminghausBrinkmann2010,RoellerBlaschkeHerminghausVollmer2014}.

The {present study} is based on extensive simulations of a two
dimensional wet granular material whose critical force for the arrest
of flow has fully been characterized in
Ref.~\cite{RahbariVollmerHerminghausBrinkmann2010}.  {In contrast to
  that previous work we address now} Lees-Edwards shear flow. We
determine the dependence of the arrest stress, $\sigma_a$, on the
packing density, $\phi$, and the {capillary energy,} $\vareps$.

The data are obtained by careful extrapolation of a sequence of
numerical measurements of the stress for larger shear rates to
vanishing $\dot\gamma$, {where we take care to base this extrapolation
  only on systems with uniform shear profiles.}  Requiring consistency
with {the setting} where the flow is driven by an external force
\cite{RahbariVollmerHerminghausBrinkmann2010}, provides a
parameter-free prediction of $\sigma_a$ that is in excellent
quantitative agreement with the numerical results.

{The paper is organized as follows: In \Sec{prediction} we revisit the
  theory for the arrest of flow
  \cite{RahbariVollmerHerminghausBrinkmann2010}, and adapt it to
  describe the arrest stress of a uniformly sheared
  system. Subsequently, in \Sec{numerical}, we describe our system,
  its equations of motion, the approach adopted to solve them
  numerically, and the data analysis. \Sec{results} comprises the
  results on the shear-rate dependence of the stress, and its
  extrapolation to zero shear rate. We will demonstrate that for
  intermediate packing densities, $\phi$, the parameter dependence of
  the arrest stress is faithfully described by the parameter free
  prediction derived in \Sec{prediction}, and trace down the
  additional physical processes causing the differences for small and
  large $\phi$. Our main results are summarized in \Sec{conclusion}.}

\section{Predicting the Arrest Stress}
\label{sec:prediction}

We consider {uniform shear in} a {two-dimensional} system in a domain
of size $L \times L$.
{By definition, the shear stress, $\sigma_{xy}$,
corresponds to a force $\sigma_{xy} L$ required to act at the
boundaries of the system in order to maintain the uniform flow.}
Following \cite{RahbariVollmerHerminghausBrinkmann2010} we identify
the critical force {persisting at very small shear rates}
based on a power balance of {the work injected into, and dissipated in
  the system.}
Energy is dissipated by breaking capillary bridges spanning the stress
network. Hence, the dissipated power takes the form
\begin{equation} 
  \langle P_{\textrm{diss}} \rangle = \int_{0}^{L} dx \; n_s(x) \;
  \left| \frac{d v_y(x)}{dx} \right| \; \nu \varepsilon = N \;
  \dot\gamma \; \nu \varepsilon
\end{equation}
where $\varepsilon$ is the 
energy needed to rupture a capillary bridge, and $\nu$ the {average}
number of bridges ruptures when two {disks} pass in the shear flow. To
arrive at {the flow configuration}
(i) the spatial density of {disks}, $n_s(x)$, per unit length is the
ratio of the total number {of disks}, $N$, and the width, $L$, of the
system, and (ii) the shear rate is constant, ${d
  v_y(x)}/{dx}=\dot\gamma=$const.

On the other hand, for a system with uniform density and shear the
total injected power is the product of the off-diagonal element of the
stress tensor, $\sigma_{xy}$, the shear rate, $\dot\gamma$, and the
system area, $L^2$, 
\begin{equation} 
  \langle P_{\textrm{forcing}} \rangle = \sigma_{xy} \, \dot\gamma \,
  L^2 \, .
\end{equation}

{Flow} ceases when {the forcing} injects too little energy to
balance dissipation.  The threshold value, $\sigma_a$, is obtained by
balancing $\langle P_{\textrm{diss}} \rangle$ and $\langle
P_{\textrm{forcing}} \rangle$,
\begin{equation} 
  \sigma_a = n_a \: \nu\varepsilon
  \label{eq:sig-a}
\end{equation} 
where $n_a=N/L^2$ is the areal density of {disks.}

\begin{figure}
  \centering
  \includegraphics[width=0.45\textwidth]{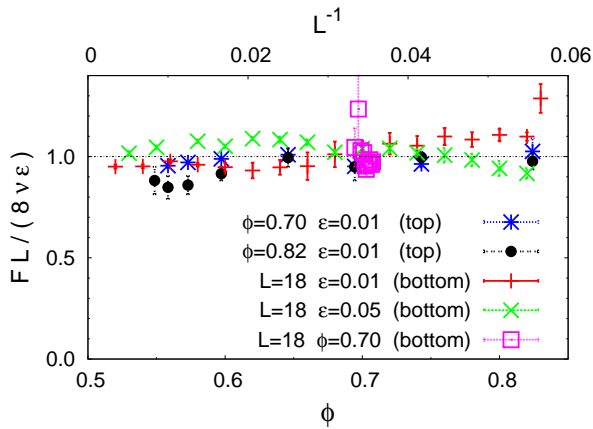}
  \caption{(color online) Fit of the amplitude, $F$, of the critical
    cosinus force field required to maintain a flow in periodic
    boundary conditions. In
    Ref.~\cite{RahbariVollmerHerminghausBrinkmann2010} it was shown
    that $F = 8 \varepsilon \nu /L$, and we show here that the
    numerical data discussed in that paper are well described by the
    expression, \Eq{nu}, for $\nu$.  The data shown by stars and
    filled circles refer to simulations at fixed $\vareps=0.01$ and
    packing densities $\phi=0.7$ and $0.82$, respectively, where we
    varied the system size $L$ as indicated by the top axis of the
    graph.  The data shown by pluses and crosses show results of
    simulations for a systems size $L=18$, capillary energies
    $\vareps=0.01$ and $0.05$, respectively, and varying densities
    $\phi$ as indicated on the bottom axis.  Finally, the open squares
    show data for $L=18$, $\phi=0.7$, and different values of
    $\vareps$ ranging between $10^{-4}$ and $10^{-1}$.
  \label{fig:cos-shear-nu-test}}
\end{figure}

In this expression the number $\nu$ depends on the packing fraction,
$\phi$, and on the capillary energy, $\varepsilon$, because collective
{motion is needed when disks} pass each other in a dense
system\@{}---\@{}as initially discussed for a system {where flow is}
driven by an external cosinus-shaped force field
\cite{RahbariVollmerHerminghausBrinkmann2010}. {In
  Ref.~\cite{RahbariVollmerHerminghausBrinkmann2010}, based on
  systematic data collapses and heuristic scaling arguments, we have
  derived an explicit analytical formula for the dependence of the
  average bridge number, $\nu$, on the packing fraction, $\phi$, and
  the capillary energy, $\vareps$. As an alternative equation, in
  \Fig{cos-shear-nu-test} we show that the data {discussed} in
  \cite{RahbariVollmerHerminghausBrinkmann2010} are also well
  described by the following equation}
\begin{equation}
  \nu = - B\;\left(\frac{\phi}{\phi_{\textrm{rcp}}-\phi}\right)^{1/2}
  \;\ln\left(\frac{4\,\varepsilon}{3}\right) \;
  \label{eq:nu}
\end{equation} 
with $B = 0.20\pm 0.02$.

In the following we explore in how far the prediction, \Eq{sig-a},
complemented with the expression, \Eq{nu}, provides a faithful
description of the arrest stress in systems where the shear rate,
$\dot\gamma$, is prescribed.

\section{Numerical Method}
\label{sec:numerical}

We numerically solve Newton's equation of motion of a 1:1 mixture of
large and small disks moving in a two-dimensional domain of size
$L\times L$ subjected to Lees-Edwards boundary conditions
\cite{Allen1987} that enforce a prescribed total shear, $L\,
\dot\gamma$, in the system. {All velocities and forces are calculated
  based on the appropriate representation of the flow in the
  two-dimensional plane that is obtained by properly unfolding the
  Lees-Edwards boundary conditions (\cf\cite{Allen1987}).} Crystalline
ordering is prevented by considering a 1:1 mixture of large and small
disks with a ratio $R_l/R_s = 1.4$ of their respective radii $R_l$ and
$R_s$~\cite{ohern2003}. The random-close packing limit of this mixture
has been reported \cite{ohern2003, Xu2005} to be $\phi_\text{rcp} =
0.84$. We assume a constant mass per area, $\rho$, such that the mass
of disk $i$ is $m_i= \pi\rho R^2_i$.

\subsection{Equations of motion}

We adopt a fifth-order predictor-corrector Gear algorithm
\cite{Allen1987,NumericalRecipes} to solve Newton's equation of motion
\begin{equation}
  m_i\frac{d^2{\bf r}_i}{dt^2} = \sum_{j\in{\cal N}(i)} {\bf
    e}_{ij}\,f_{ij}(r_{ij})~,
  \label{eq:Newton}
\end{equation}
where $m_i$ and ${\bf r}_i$ are the mass and the center disk $i$,
respectively, ${\cal N}(i)$ is the set of neighbors $j$ interacting
with $i$, the unit vector ${\bf e}_{ij}$ points from the center of
disk $i$ to the center of disk $j$, and $f_{ij}$ is the force exerted
by disk $i$ on disk $j$. The latter force comprises a repulsion, and
whenever applicable also the attractive force, \Fcapi, modeling
capillary bridges.

Mutual repulsion between the disks is modeled by the repulsive force
\begin{subequations}
  \begin{equation} 
    F_r(r_{ij}) = \left\{
    \begin{array}{ll}
      C_{ij}(R_i+R_j-r_{ij})^{1/2} &  \mbox{for } r_{ij} \le R_i+R_j \\
      0                            &  \mbox{else,} 
    \end{array}
    \right. 
  \end{equation} 
  where $r_{ij}$ is the Euclidean distance between he center of disk
  $i$ and $j$. In the spirit of Hertz's contact law
  \cite{poeschel2005} we set
  \begin{equation}
    C_{ij} 
    = C 
    \left(
    \frac{R_i R_j}{R_i + R_j}
    \right)^{1/2}
  \end{equation}%
  \label{eq:non_linear_spring_force}%
\end{subequations}
in order to account for different disk radii. The global parameter $C$
controls the hardness of the disks.

For the capillary bridges forces we engage the minimal capillary model
proposed by Herminghaus in Ref.~\cite{herminghaus2005} in which the
capillary force is constant, $\Fcapi$, as the distance between
granulates changes, and where it breaks at a distance $s_c$. This
model has been shown to be well-suited for numerical simulations of
wet granular materials~\cite{RoellerBlaschkeHerminghausVollmer2014,
  fingerle2006,fingerle2008, herminghaus2004}. The force in this model
is hysteretic: upon first approach there is no attractive force
{($r_{ij} > R_i + R_j$)}; when the disks undergo collision
{($r_{ij} \le R_i + R_j$)} a constant (capillary) force $\Fcapi$ is
activated, and the force persists until the disks separate by a
(surface-to-surface) distance $s_c$ {($r_{ij} \le R_i + R_j +
  s_c$)}.  At that point, {$r_{ij} = R_i + R_j + s_c$}, the
capillary bridge is removed (it `ruptures'). There is no force acting
between the disks again, until they undergo their next collision
{($r_{ij} \le R_i + R_j$)}. Consequently, an energy $\epsilon =
\Fcapi s_c$ is dissipated {after rupture of the capillary bridge.}

Throughout this paper, we employ dimensionless rescaled quantities
based on the capillary force, \Fcapi, the mass density of the disks,
$\rho$, and the average disk diameter $D$. Time, $t$, and mass, $m$,
is hence measured in units $\tau \equiv \sqrt{\rho D^3/\Fcapi}$, and
$\mu \equiv \rho D^2$, respectively.  Using these normalized
quantities it is straightforward to normalize all physical quantities
derived from mass, length and time, such as the local averages of disk
velocities, {components of the stress tensor,} and the shear rate.

\subsection{Data acquisition and analysis}

For each parameter set we run $20$ simulations over a fixed total
strain of $\gamma_t = 10\times L$. The choice to fix the total strain,
rather than fixing the total simulation time, ensures that in the
limit of very small flow rates we still sample a representative set of
statistically uncorrelated disk configurations. We start to measure
the physical quantities at the strain $\gamma = 4 \times L$ and take
$150$ shots over the rest of the simulations until $\gamma_t =
10\times L$.  Altogether these $150$ configurations for $20$ different
runs provide $3000$ independent snapshots for each considered
parameter set. For each snapshot, we determine and store the
components of the stress tensor, ${\bf \sigma}$, as well as the
average density, partial densities, and granular temperature evaluated
in ten bands parallel to the shear.

The components of the stress tensor are
calculated~\cite{lois2005, RahbariVollmerHerminghausBrinkmann2010,rahbari2013}
by evaluating

\begin{eqnarray}
  \sigma_{\alpha\beta} &=& \frac{1}{L^2} \; \sum_i m_i \: \left(
        {v}_{i,\alpha} - {U}_{\alpha}({\bf x}_{i}) \right) \: \left(
        {v}_{i,\beta } - {U}_{\beta }({\bf x}_{i}) \right)
        \nonumber\\ &+& \frac{1}{L^2} \; \sum_{i<j} \: {r}_{ij,\alpha}
        \: {F}_{ij,\beta} \, .
        \label{eq:stress_tensor}
\end{eqnarray}
In \Eq{stress_tensor} $\alpha,\beta \in \{ x,y \}$ denote the
Cartesian components of the respective vectors or tensors. The stress
tensor has two contributions: (I) the first term accounts for
{advective} momentum transport, where , ${\vec v}_{i}$ is the velocity
of the disk $i$, and ${\vec U}({\vec x}_i)$ is the local drift
velocity at the position ${\vec x}_i$ of disk $i$, \ie the overall
center-of-mass velocity of the disks with centers in an interval of
width $L/10$ containing $x_i$.  (II) the second term describes the
contribution of the interaction forces between particles on the stress
{to the flux of linear momentum}. In this contribution ${\bf r}_{ij}$
is the vector connecting the centers of disk $i$ and $j$, and ${\bf
  F}_{ij}$ is the force exerted by the disk $i$ on disk~$j$. In the
limit of small shear rates, $\dot\gamma \rightarrow 0$, one approaches
creep flow, and the latter term in Eq.~\ref{eq:stress_tensor} always
dominates the shear stresses.

\hb2{ For non-Newtonian fluids, the relationship between the shear
  stress, $\sigma_{xy}$, and the shear rate, $\dot\gamma$, is given by
  Herschel-Bulkley relation~\cite{herschel_1926}:}
\begin{equation}
  \sigma_{xy} = \sigma_a + K \times \dot{\gamma}^n
  \label{eq:herschel}
\end{equation}
  where, $\sigma_a$, is the arrest (yield) stress, $K$, is a constant,
  and, $n$, is the shear thinning exponent. \hb2{According to a recent
    review by Bonn {\it et al.}~\cite{bonn_2015}, the shear thinning
    exponent is a material dependent parameter for which no
    universality has been found. Therefore, Eq.~\ref{eq:herschel}
    would be a natural choice to be considered as a template function
    for experimental data fitting. However, we found out that when the
    shear thinning exponent, $n$, is variable, for large capillary
    energies, $\vareps>0.05$, the fitting algorithm frequently takes
    very small exponents in the order of $10^{-1}$. This very small
    exponent gives rise to a flow profile which is very steep near the
    origin. As a result of the sharp fall near the origin, it becomes
    more likely that the fitting algorithm grasps a negative
    offset. The negative offset corresponds to a negative arrest
    (yield) stress, $\sigma_a$, which is non-physical. In order to
    prevent this fitting failure, it is required that the final
    fitting flow profile should be smooth at the origin. This is
    guaranteed by enforcing $n>1$. To be consistent, we choose a
    quadratic profile, $n=2$, which is the most trivial choice given
    by the Bagnold scaling~\cite{VoagbergOlssonTeitel2013,
      IraniChaudhuriHeussinger2014}.}

\begin{figure}
  \centering
  \includegraphics[width=0.5\textwidth]{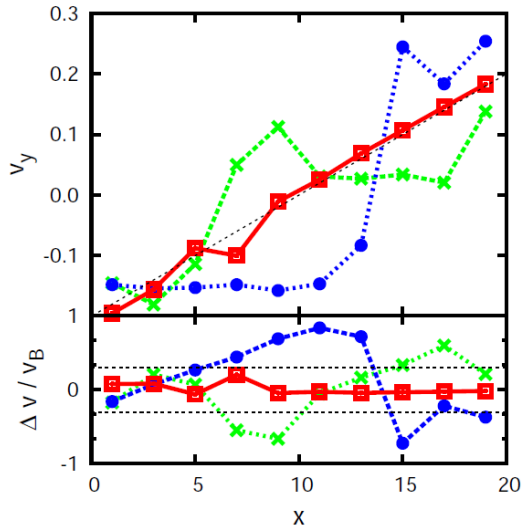}
  \caption{(color online) The upper panel shows the mean
    center-of-mass velocity in ten stripes parallel to the flow for
    three shear profiles encountered in simulations for shear rate
    $\dot{\gamma} = 0.02$, system size $L=20$, packing fraction
    $\phi=0.82$, and capillary-bridge energy $\vareps=0.1$.  The lower
    panel shows the respective relative deviations from the ideal
    profile. The open red squares connected by a solid line mark a
    profile where the deviations from the mean profile are in the
    admissible bounds of $\pm 30$\% that are indicated by thin black
    lines.  Green crosses and blue circles, that are connected by
    broken lines in according color, show profiles where the system
    has formed a shear band.
    \label{fig:shearbands}}
\end{figure}

\subsection{Uniform shear profiles}

In the analysis {of the arrest stress} we systematically disregard all
the snapshots where the velocity profile deviates by more than a
factor of $0.3$ from the ideal linear case
\begin{equation}
  v_y(x) = \dot{\gamma} \; \left( x - \frac{L}{2} \right) \, ,
  \label{eq:ideal_Vy}
\end{equation}
hence excluding phase-separated states (\cf\Fig{shearbands}). { The
  ratio at which the snapshots are disregarded from further analysis
  strongly depends on the packing fraction, $\phi$. The higher the
  packing fraction, the lower the exclusion ratio. As an example, for
  $\phi = 0.52$, the rejection ratio is approximately $\simeq 72\%$,
  and for $\phi = 0.83$, it is about $\simeq 3\%$.}

{The lower panel of \Fig{shearbands} shows the respective relative
  deviations from the ideal profile in order to exemplify the
  criterion for the selection of valid flow configurations.  Flow
  configuration with deviations of the average flow velocities from
  the ideal profile (the thin dotted line in the upper panel of the
  figure) up to $\pm 30$\% commonly appear as fluctuations on
  statistically uniform shear profiles: these flow configurations will
  enter the analysis of the arrest stress.  On the other hand, larger
  fluctuations rapidly evolve into shear bands that are persistent in
  time\@{}---\@{}once they appear, they do not decay again. The green
  crosses and blue circles, that are connected by broken lines in
  according color in \Fig{shearbands}, show profiles where the system
  has formed a shear band. In that case the deviations lie noticeably
  outside the admissible range, and they persist: the system has phase
  separated into two domains of almost constant drift velocity;
  localizing the shear in narrow bands that lie at $5 \simeq x \simeq
  10$ (green crosses) and $12 \simeq x \simeq 15$ (blue circles),
  respectively.  These configurations must be excluded from the
  analysis because they show a region with a very low (often zero)
  number of capillary bridges, and hence anomalously low shear
  stress.}

\section{Results {and Discussion}}
\label{sec:results}

\begin{figure}
  \[
  \includegraphics[width=0.45\textwidth]{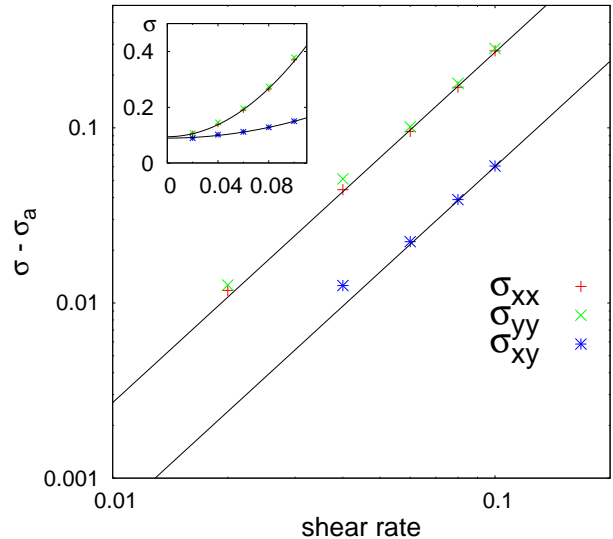}
  \]
  \caption{{(Inset) The shear rate, $\dot\gamma$, dependence of the
      components of the stress tensor, $\sigma$. The red pluses, green
      crosses, and blue stars show $\sigma_{xx}$, $\sigma_{yy}$, and
      $\sigma_{xy} = \sigma_{yx}$, respectively. The solid lines show
      the best fits by a quadratic function,
      $\sigma_{\alpha\beta}(\dot\gamma) = a_{\alpha\beta} +
      b_{\alpha\beta} \dot{\gamma}^2 $, to the data where,
      $\sigma_{xx} = 0.095 + 27 \dot\gamma^2$, $\sigma_{yy} = 0.097 +
      27 \dot\gamma^2$, and $\sigma_{xy} = 0.09 + 6 \dot\gamma^2$. One
      can see that all the components of the stress tensor possess
      finite offsets at $\dot\gamma=0$. This refers to a finite
      dynamic yield stress of wet granular materials for $\phi <
      \phi_{rcp}$. (Main Panel) Components of the stress tensor
      subtracted from their corresponding offsets, $\sigma -
      \sigma_a$, in a double logarithmic scale. In these simulations,
      the packing fraction is $\phi = 0.78$, capillary energy is
      $\vareps = 0.05$, and box length is $L = 20$.}
    \label{fig:converg}}
\end{figure}


\subsection{Shear-rate dependence of the arrest stress}

{In \Fig{converg}-Inset, we show the dependence of the components of
  the stress tensor on the shear rate $\dot{\gamma}$.} The diagonal
elements of the stress tensor, $\sigma_{xx}$ and $\sigma_{yy}$, are
depicted by {red pluses and green crosses, respectively, and the blue
  stars} show the off-diagonal element $\sigma_{xy} =
\sigma_{yx}$. All curves {show} finite offsets at $\dot{\gamma} = 0$.
The {solid lines show the best fits by quadratic functions
  $\sigma_{\alpha\beta}(\dot\gamma) = a_{\alpha\beta} +
  b_{\alpha\beta} \dot{\gamma}^2$, where $a_{\alpha\beta}$ accounts
  for the non-vanishing stress at zero shear rates, and
  $b_{\alpha\beta} \dot{\gamma}^2$ is the Bagnold scaling expected to
  arise at large $\dot\gamma$} \cite{VoagbergOlssonTeitel2013,
  IraniChaudhuriHeussinger2014}. In the main {panel} of \Fig{converg},
we verify this quadratic scaling for large $\dot\gamma$ by plotting
the components of $\sigma-a$ on a double logarithmic scale.  The
offset of the off-diagonal element provides the arrest stress,
$\sigma_a=a_{xy}=a_{yx}$. It is an intensive parameter of the system
that does not depend on the system size. {However, clearly, the
  components of $a$ and $b$ are functions of the density, $\phi$, and
  the capillary energy, $\vareps$.}

\subsection{Parameter dependence of $\sigma_a$}

\begin{figure}
  \centering
  \includegraphics[width=0.5\textwidth]{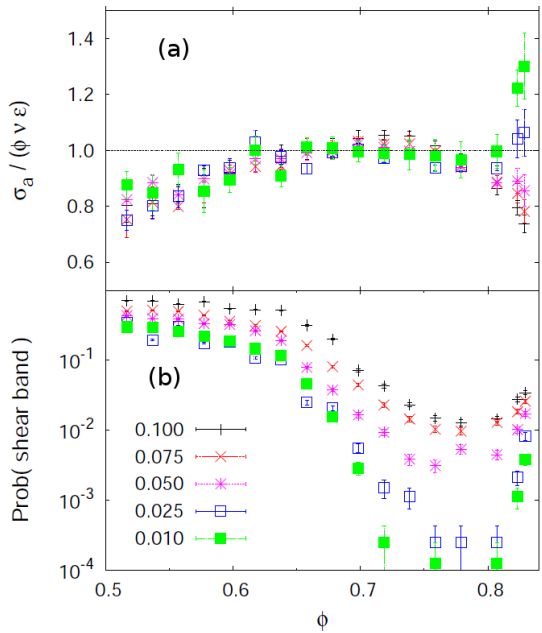}
  \caption{{\bf (a)} collapse of the data according to \Eq{sig-a},
    which asserts that $\sigma_a /(\phi \nu \vareps)$ should be one
    (as indicated by the dotted line), as function of the packing
    fraction, $\phi$, for $L=20$ and five different capillary
    energies, $\vareps$, that are provided in the lower left corner of
    the panel-b. {\bf (b)} The panel provides the probability to
    observe shear bands in the simulations. {In the legend the
      corresponding capillary energy, $\vareps$, of each simulation is
      given}.}
    \label{fig:vonMises_density}
\end{figure}

\Fig{vonMises_density}-a demonstrates our data collapse according to
\Eq{sig-a} for {the dependence of} the arrest stress, $\sigma_a$,
{on} the packing fraction, $\phi$, for five different capillary
energies, $\vareps$. For $0.6 \lesssim \phi \lesssim 0.8$ we find an
excellent agreement between our data and the prediction
$\sigma_a/(\phi\nu\vareps) = 1$ shown by the dotted line.

The data collapse is breaking down upon approaching the random close
packing limit for values $\phi \gtrsim 0.8$.  We expect that in this
range the repulsive forces gives rise to an additional contribution to
the stress, as observed in \cite{IraniChaudhuriHeussinger2014}.

For $\phi \lesssim 0.6$ the values of $\sigma_a$ systematically fall
below the prediction, \Eq{sig-a}.  This systematic error results from
the choice $\sigma_a + b_a \dot\gamma^2$ of the function selected for
extrapolating the $\dot\gamma$ dependence of $\sigma_{xy}$ to small
$\dot\gamma$: it does not account for the slight decrease of
$\sigma_{xy}$ for small values of $\dot\gamma$ in the regime where one
expects shear banding (\cf for instance Fig.~1 of
\cite{IraniChaudhuriHeussinger2014}).  As shown in the main panel of
\Fig{converg} there always is a small mismatch of the simulation data
and the fit function for small $\dot\gamma$. We still adopted the
function $a + b \dot\gamma^2$ because it involves fewer parameters and
results in a more robust fitting of our data.  To within our numerical
error margins the systematic error in the fit is negligible for
$\phi\gtrsim 0.6$, where the dip becomes so minute that shear banding
is no longer encountered.  On the other hand, for $\phi\lesssim 0.6$
the fit provides a value characterizing the minimum of
$\sigma_{xy}(\dot\gamma)$ rather than the asymptotic value for
vanishing $\dot\gamma$. When this distinction becomes noticeable the
system becomes prone to shear banding ( \Fig{vonMises_density}-b).

\section{Conclusions}
\label{sec:conclusion}

We calculated the arrest stress, $\sigma_a$, characterizing the
minimal value of $\sigma_{xy}(\dot\gamma)$ for two-dimensional
isotropic shear flows of wet granular disks by fitting the components
of the stress tensor by a quadratic functions, $a+b \dot\gamma^2$.  In
the range $0.52 < \phi < 0.84$, all components of the stress tensor of
wet granular materials approach non-trivial finite values for small
$\dot\gamma$. The resulting offset value for the shear stress is of
particular interest.  In general it is different from the {static
  yield stress}, $\sigma_y$, that can be obtained
\cite{IraniChaudhuriHeussinger2014} by considering the limit $\sigma_y
= \lim_{\dot\gamma\to 0} \sigma_{xy}$.  The latter characterizes the
stress where a system at rest can start to flow. {In the case of
  wet granular materials, the static yield stress is the force which
  is needed to break down the percolation network of capillary
  bridges}. The former, {\ie the dynamic yield stress}, is the minimal
value of the stress which is required to maintain the flow.


The prediction of the arrest stress is based on a balance of the power
injected by increasing strain at a finite stress and the energy
dissipation due to rupturing capillary bridges. For uniform shear this
argument provides \Eq{sig-a}. This prediction has a single free
parameter, the average number, $\nu$, of bridges ruptures when
particles pass each other in the flow. We have determined this number
\cite{RahbariVollmerHerminghausBrinkmann2010} in the context of a
different flow, non-uniform shear flow driven by an external force
field in periodic boundary conditions.  Evoking universality of the
number, $\nu$, with respect to the cause of the flow provides a
prediction of the arrest stress without adjustable parameters. The
excellent data collapse shown in \Fig{vonMises_density}-a confirms the
versatility of this approach to determine the arrest of flow
\cite{RahbariVollmerHerminghausBrinkmann2010,rahbari2013,RoellerBlaschkeHerminghausVollmer2014},
and the universality of $\nu$ as far the protocol of inducing the flow
are concerned. The difference between our prediction, \Eq{sig-a}, and
the numerical findings for $\phi \gtrsim 0.8$ have been associated to
the emergence of contributions to the stress from repulsive forces
when approaching random close packing.  The systematic difference for
$\phi \lesssim 0.6$ is due to a noticeable deviation of the arrest
stress from {the static yield stress}. The larger the difference, the
more likely the flow will phase separate into a state featuring shear
bands (\cf \Fig{vonMises_density}-b).

{Previous studies of far-from equilibrium transitions in wet granular
  materials have unraveled origins of these transitions to be either
  force- or energy-driven~\cite{strauch2012, fingerle2008}. For
  instance, for wet granulates in a Petri dish under vertical
  vibration, solid-to-fluid (fluid-to-gas) transition is force-driven
  (energy-driven). In sheared wet granular matter, the solid-to-fluid
  transition, whose threshold represents the static yield stress, has
  been shown to be of force-driven
  origin~\cite{RahbariVollmerHerminghausBrinkmann2009}}

 The present work complements and further substantiates the findings
 of previous studies
 \cite{RahbariVollmerHerminghausBrinkmann2010,rahbari2013,RoellerBlaschkeHerminghausVollmer2014}
 that attributed the fluid-to-solid transition in sheared wet granular
 systems is connected to a balance of the power injected by the
 forcing of the flow and the energy dissipation rate due to rupturing
 of capillary bridges in a plastic shear flow
 ({energy-driven}). This mechanism is fundamentally connected to
 the fact that wet granular materials have an inherent energy scale,
 the energy $\vareps$ that is dissipated upon rupturing a capillary
 bridge \cite{herminghaus2004,herminghaus2005}. It will be interesting
 to explore the impact of this important distinction for other soft
 matter systems
 \cite{LoisBlawzdziewiczOHernCorey2008,ChaudhuriBerthierBocquet2012,IraniChaudhuriHeussinger2014}
 where small attractive interactions and a dissipation due to transfer
 of kinetic energy into internal degrees of freedom of the system
 should admit a corresponding analysis of the arrest of flow.

This is in contrast to dry granular materials where the arrest of flow
is expected to be controlled by frictional forces between the
particles.

\begin{acknowledgments}

We would like to thank M. Akbari-Moghanjoughi for his
comments. S.H.E.R. is financially supported by the Iran National
Science Foundation (INSF) Grant No. 90004064.

\end{acknowledgments}




\end{document}